\documentclass[fleqn,10pt]{wlscirep}

\usepackage[utf8]{inputenc}
\usepackage[T1]{fontenc}
\usepackage{color}

\title{Direct test of the FLRW metric from strongly lensed gravitational wave observations}

\author[1]{Shuo Cao}
\author[2]{Jingzhao Qi}
\author[1,+]{Zhoujian Cao}
\author[1,3]{Marek Biesiada}
\author[4]{Jin Li}
\author[5]{Yu Pan}
\author[1,*]{Zong-Hong Zhu}
\affil[1]{Department of Astronomy, Beijing Normal University,
Beijing 100875, China}
\affil[2]{Department of Physics, College of Sciences,
Northeastern University, 110819 Shenyang, China}
\affil[3]{Department of Astrophysics and
Cosmology, Institute of Physics, University of Silesia,
75 Pu{\l}ku Piechoty 1, 41-500, Chorz{\'o}w, Poland}
\affil[4]{Department of Physics, Chongqing University, Chongqing
400030, China}
\affil[5]{College of Science, Chongqing University of Posts and
Telecommunications, Chongqing 400065, China}

\affil[*]{zhuzh@bnu.edu.cn}

\affil[+]{zjcao@amt.ac.cn}

\begin{abstract}
The assumptions of large-scale homogeneity and isotropy underly the
familiar Friedmann-Lema\^{\i}tre-Robertson-Walker (FLRW) metric that
appears to be an accurate description of our Universe. In this
paper, we propose a new strategy of testing the validity of the
FLRW metric, based on the galactic-scale lensing systems where
strongly lensed gravitational waves and their electromagnetic
counterparts can be simultaneously detected. Each strong lensing
system creates opportunity to infer the curvature parameter of the
Universe. Consequently, combined analysis of many such systems will
provide a model-independent tool to test the validity of the FLRW
metric. Our study demonstrates that the third-generation ground
based GW detectors, like the Einstein Telescope (ET) and space-based
detectors, like the Big Bang Observer (BBO), are promising
concerning determination of the curvature parameter or possible
detection of deviation from the FLRW metric. Such accurate
measurements of the FLRW metric can become a milestone in precision
GW cosmology.
\end{abstract}

\begin{document}

\flushbottom
\maketitle
%
%

\section*{Introduction}

It is well known that Friedmann--Lema\^{\i}tre--Robertson--Walker
(FLRW) metric, an exact solution of the Einstein's equations
obtained under the assumption of homogeneity and isotropy of space,
is very successful in explaining many observational facts concerning
our Universe including large-scale distribution of galaxies and the
near-uniformity of the CMB temperature \cite{Ade16}. A particularly
successful application of the FLRW metric is that it underlies the
present standard cosmological model, which is very successful in
fitting the current observational data sets and explaining the
observed cosmic acceleration. Significant evidence that the
space-time metric in cosmological scales deviates from the FLRW
metric would have profound consequences for inflation theory and
fundamental physics. For instance, the observed phenomenon of
late-time accelerated expansion can be attributed to the failure of
the FLRW approximation \cite{Rasanen04,Kolb05}, which motivated the
search for alternative solutions including the effect of
inhomogeneities in small scales on the average expansion rate
\cite{Boehm13,Lavinto14}, as well as the violation of statistical
homogeneity and isotropy
\cite{Ellis05,Rasanen11,Buchert12,Redlich14,Montanaria17}. The last
two decades brought rapid advances in observational cosmology, which
made it possible to test FLRW metric by checking the consistency
between different observables derived from general assumptions of
geometrical optics \cite{Clarkson08,Shafieloo10,Sapone14,Rasanen14}.
Recently, \cite{Rasanen15} presented a new test of the validity of
the FLRW metric, using the combination of independent observations:
Union2.1 supernova distances and galaxy strong lensing data from
Sloan Lens ACS Survey. Let us note that the previous works focused
on the reconstruction of a smooth distance function with supernova
data, which were used to estimate the luminosity distance to lenses
and sources \cite{Cai15,Denissenya18}.

On the other hand, the first direct detection \cite{Abbott16a} of
the gravitational wave (GW) source GW150914  has opened an era of
gravitational wave astronomy and added a new dimension to the
multi-messenger astrophysics. Possibility of using GW signal from
inspiraling binary system to determine the absolute value of its
luminosity distance and thus the Hubble constant were first noticed
in  \cite{Schutz86}. In recent works inspiraling binary black holes
(BHs) and neutron stars (NS) have successfully been used as standard
sirens \cite{Abbott17c}. Moreover, the effect of gravitational
lensing of GWs has recently gained interest and has been studied
extensively. The earliest attempts to discuss lensing of GW signals
from merging NS binaries, its influence on the inferred luminosity
distance to the source, expected total number of events and their
redshift distribution can be traced back to \cite{Wang96}. Some
recent works took even more extreme view suggesting that the first
six binary BH merging events reported by LIGO/Virgo should be
reinterpreted as being lensed and proposing that GW170809 and
GW170814 events are the lensed signals from the single source
\cite{Broadhurst19}. However, The interest in detecting a lensed GW
signal is powered by suggestions that cosmological parameters can be
significantly constrained using time delays measurements of strongly
lensed GW events \cite{Liaokai2017}. Moreover, strongly lensed GW
signals can be used to test fundamental physics. For example, the
speed of gravity can be tested with strongly lensed GW events
accompanied by electromagnetic counterparts \cite{Fan2017}. Recent
analysis revealed that detection of lensed events with advanced LIGO
detectors could be quite plausible \cite{Li18}. In the context of
third generation interferometric detectors, a series of papers
\cite{Ding15} explored the perspectives of observing gravitationally
lensed coalescing double compact objects in the Einstein Telescope
(ET). The detailed calculation of GW lensing rate caused by lensing
galaxies showed that ET would register about 100 strongly lensed
inspiral events per year at its design sensitivity.
In this paper, we propose that lensed GW signals accompanied by EM
counterpart would enable determination of cosmic curvature and would
provide a test of validity of the FLRW metric. The main motivation
of this work is that lensed GWs can be probed at much higher
redshifts than SNIa or the large scale structure \cite{Qi18a}.

Homogeneous and isotropic spacetime is described by the FLRW metric
\begin{equation}
ds^2= c^2dt^2 - \frac{a(t)^2}{1-Kr^2}dr^2 - a(t)^2r^2d\Omega^2,
\end{equation}
where $a$ represents the scale factor (in the unit of [length]). The
dimensionless curvature $K$ is related to the cosmic curvature
parameter as $\Omega_k=-K c^2 /a_0^2 H_0^2$, where $H_0$ denotes the
Hubble constant (in the unit of [time]$^{-1}$) and $a_0$ is the
present value of the scale factor. Let us introduce dimensionless
comoving distances $d_{ls}\equiv d(z_l,z_s)$, $d_l\equiv d(0,z_l)$
and $d_s\equiv d(0,z_s)$, related to the (dimensioned) comoving
distances $D$ as $d=H_0D/c$. In the subsequent simulations of
strongly lensed GW data, we use the Hubble constant measurement of
$H_{0}=67.8\pm0.9$ km $\rm s^{-1}$ $\rm Mpc^{-1}$, determined from
Planck 2015 temperature data combined with Planck lensing
\cite{Ade16}. In the flat universe, dimensionless comoving distances
between the observer and two other aligned objects $l$ and $s$ will
obey an additivity relation $d_{ls} = d_{s} - d_{l}$. In non-flat
FLRW models this relation is modified to $d_{ls}
=\sqrt{1+\Omega_kd_l^2} d_s - \sqrt{1+\Omega_k d_s^2} d_l$
\cite{Rasanen15}. Such setting is realized in Nature by strong
gravitational lensing systems, where the source located at redshift
$z_s$ and intervening galaxy acting as a lens located at redshift
$z_l$ are almost perfectly aligned. Admittedly, we use FLRW metric
in order to be specific how to calculate distances in the distance
sum rule. However, in our method all necessary distances can be
assessed in each strong lensing system directly without tacit
assumptions underlying the extrapolation from SN Ia, for example.
FLRW metric invoked is just a framework whose consistency is being
checked. From the location of multiple images of a strong lensing
system, one is able to assess the ratio of angular-diameter
distances $ \frac{D^{A}_{ls}}{D^{A}_s} = \frac{d_{ls}}{d_s}$ (see
the Methods section for details). Therefore, by virtue of the
distance sum rule \cite{Rasanen15}, one is able to measure
\begin{equation} \label{smr}
\frac{d_{ls}}{d_s}=\sqrt{1+\Omega_kd_l^2}-\frac{d_l}{d_s}\sqrt{1+\Omega_k
d_s^2}.
\end{equation}
If the other two dimensionless comoving distances $d_l$ and $d_s$
can be measured, then the value of $\Omega_k$ could be determined.
As we will discuss in more details in the Methods section, all these
necessary ingredients can be derived from multiple information
accessible in particular type of lensing systems. In this paper, we
focus on galactic-scale strong gravitational lensing systems with
high-redshift inspiraling NS-NS and NS-BH binaries acting as
background sources. The EM counterpart would allow the host galaxy
and lens galaxy identification and state-of-the-art lens modelling
techniques would enable a precise reconstruction of lens mass
distribution.

One of the typical features of lensed GW signals is that time delays
between lensed images (e.g. 1$\sim$100 days) inferred from the GW
observations would have uncertainties (e.g. $\sim 0.1\;s$ from the
detection pipeline) totally negligible comparing to the uncertainty
in lens modeling, while other relevant observables, like redshifts
or images, can be precisely measured in the EM domain. This allows
the assessment of the time delay distance $D_{\Delta t}$ and the
transverse comoving distance to the lens $D_l$ (see the Methods
section for details).

Finally, as already mentioned, GW signals from inspiraling and
merging compact binary systems can provide the luminosity distance
from the observer to the source $D^L_s$. The lensing amplification
$\mu$ boosts\cite{Wang96} the amplitude of the GW strain signal $A$
by a factor $\sqrt{\mu}$ and accompanying EM flux by a factor $\mu$.
Once the lens potential is recovered from the image analysis
combined with time delays, the magnification factors of images can
be estimated. Consequently, the transverse comoving distance $D_s$
can be derived. Proceeding along the steps outlined in the Method
section, we find that the $\Omega_k(z_l,z_s)$ values for different
pairs $(z_l, z_s)$ can be directly obtained. Due to the strong
covariance between $d_l$, $d_s$, and $d_{ls}$, instead of
propagating distance uncertainties, we use Monte-Carlo simulation to
project uncertainties in the lens mass profile, time delays, Fermat
potential difference, and the magnification effect onto the final
uncertainty of $\Omega_k (z_l, z_s)$. This creates an opportunity to
test the validity of the FLRW metric. Namely, in the FLRW universe
$\Omega_k(z_l,z_s)$ should be constant and equal to the present
value of curvature parameter e.g. inferred from CMB anisotropies.
However, if one notices significant differences between
$\Omega_k(z_l,z_s)$ values for different pairs $(z_l, z_s)$, which
cannot be accounted for by systematics and scatter, then this could
be a signal that FLRW description breaks down. The converse is not
true: if $\Omega_k(z_l,z_s)$ is constant, it does not indicate that
light propagation on large scales is not described by the FLRW
metric \cite{Rasanen15}.

\section*{Results}
Table I summarizes the uncertainties of multiple measurements in GW
and EM lensing systems. The justification for the values reported is
given below.

\subsection*{Precision of lens reconstruction}
Newly developed state-of-the-art lens modeling techniques and
kinematic modeling methods have demonstrated their power to extract
the information about lens mass distribution from high-quality
imaging of strong lensing systems \cite{Suyu12b}. On the other hand,
spectroscopic data for the central parts of lensing galaxies became
available, which made it possible to assess central velocity
dispersions inside the aperture. We assumed the uncertainty of the
velocity dispersion at the level of 5\% and the uncertainty of the
Einstein radius measurements at the level of 1\%. The assumed
accuracy of the Einstein radius measurements is reasonable for the
future LSST survey. One can expect that the high resolution imaging
of LSST, with different stacking strategies for combining multiple
exposures, making possible that a deeper stacked image could be
obtained with the combination of individual exposures for each
object \cite{Collett16}, will enable such precise Einstein radius
measurements. The joint gravitational lensing and stellar-dynamical
analysis of the SL2S lens sample \cite{Ruff11} has shown that it is
feasible to determine the total mass-density slope inside the
Einstein radius at the level of 5\% . However, the supplementary
information about time delays can reduce this uncertainty to 1\%
level as suggested by previous analysis of lensed quasars
\cite{Wucknitz04}. One can be optimistic in this respect,
considering over a decade of efforts made by the H0LiCOW
collaboration \cite{Suyu17} to develop techniques and gather data
with sufficient constraining power, using time delays measured by
the COSMOGRAIL collaboration \cite{Bonvin18}.

\subsection*{Precision of time-delays}

Precise time delays are the crucial point of our idea. As time
delays $\Delta t$ between strongly lensed gravitational wave signals
can be measured with an unprecedented accuracy of $\sim 0.1\;s$
 from the detection pipeline \cite{Liaokai2017} or even by many
orders of magnitude higher if the details of the waveform are
analyzed, e.g. the moment of the final coalescence can be determined
with $\sim 10^{-4} \; ms$ accuracy. Let us stress, that the Fermat
potential is of particular relevance for our analysis, due to its
direct dependence on the logarithmic slope of the mass profile at
the Einstein radius. Therefore, the uncertainty of the Fermat
potential difference should be simulated from the lens mass profile
and the Einstein radius uncertainties \cite{Suyu17}. If the EM
counterpart is detectable allowing for host galaxy image analysis
and the measurement of central velocity dispersion of the lens, one
can achieve the $\sim 0.5\%$ precision level of the Fermat potential
difference provided the lensed host image quality is typical to the
HST observations \cite{Liaokai2017}. Finally, the mass distribution
along the line of sight, i.e., LOS contamination might introduce 1\%
uncertainty of the lens potential recovery. This justifies the
assumption of a few percent uncertainty on the Fermat potential
measurements.

In order to demonstrate the performance of our method, we simulated
a population of 10000 realistic strong lensing systems that could be
observed by the Large Synoptic Survey Telescope (LSST)
\cite{Collett15}, in which the lenses are modeled with the power-law
mass distribution ($\rho \sim r^{- \gamma}$). Details of the
simulation are given in the Methods section. The main target of this
work are lensed GW signals from NS-NS and BH-NS systems with high
signal to noise ratio (SNR) and up to high redshifts. We performed
Monte Carlo simulation to create a mock catalog of luminosity
distances to the GW events that can be detected with $SNR>8$ (for
details see the Methods section). In particular, we refer to the
third generation of interferometers -- the "xylophone" configuration
for the Einstein Telescope (ET) \cite{Taylor12} and the second
generation technology for the Big Bang Observer (BBO)
\cite{Cutler09}. If the EM counterpart like a short gamma ray burst
(SGRB) could be visible, this would facilitate identification of the
host galaxy and measurement of the redshift. In addition, BBO's
angular resolution would be sufficient to uniquely identify the host
galaxy for the majority of binaries, the redshifts of which could be
obtained from a coordinated optical/infrared observing campaigns.

\begin{figure*}
\includegraphics[scale=0.5]{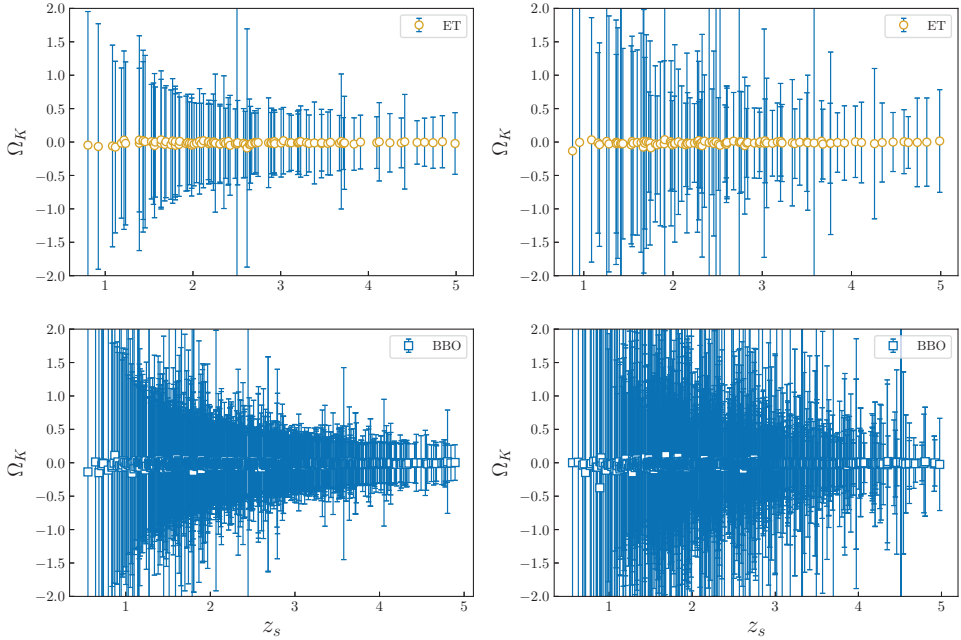}
\caption{An example of the simulated measurements of the cosmic
curvature from future observations of lensed GWs. We simulated 100
lensed GW signals detectable by the ET (upper panel) and 1000
signals detectable by the BBO (lower panel). Left and right panel
respectively show the results with 10\% and 50\% uncertainty in the
amplification factor measurements. }\label{fig1}
\end{figure*}

An example of the simulation results obtained from future ET and BBO
survey is shown in Fig.~\ref{fig1}, while the statistical
constraints on the constant FLRW parameter $\Omega_k$ are presented
in Fig.~\ref{fig2}.

\section*{Discussion}

In light of the results presented above, the question arises:
\emph{Are these measurements sufficient enough to detect possible
deviation from the Friedmann-Lema\^{\i}tre-Robertson-Walker metric?}
Considering relatively poor-precision measurements of $\Omega_k(z_l,
z_s)$ in the lower redshift region ($z<1$), it is very difficult to
achieve results competitive with other probes. However, the
distributions of lensed GWs detectable by ET and BBO have tails
reaching even $z=5$. Therefore, we expect that ET with its potential
of discovering a large number of lensed GWs will yield hundreds of
measurements of cosmic curvature. Compared with ground-based GW
detectors, space-borne detectors would provide order of magnitude
bigger catalog of more precise $\Omega_k$ measurements. This would
increase chance of finding significantly different $\Omega_k(z_l,
z_s)$  for different pairs $(z_l, z_s)$ in the case when FLRW metric
breaks down on some large scale. In principle, the function
$\Omega_k(z_l, z_s)$ can be reconstructed from observations, and the
FLRW metric is ruled out if $\Omega_k(z_l, z_s)$ is not constant
\cite{Clarkson08,Shafieloo10,Sapone14}. However, one should note
that, like all nonlinear functions of the measurements, the
definition of $\Omega_k(z_l, z_s)$ provided in Eq.~(\ref{k}) could
provide a biased estimate of cosmic curvature due to the nonlinear
propagation of measurement uncertainties. In any case, given the
limited sample size of lensed GW events, we do not try to find
$\Omega_k$ as a function of redshifts. Instead, we fit a constant
$\Omega_k$ to the data and consider the accuracy of fit. The problem
was previously recognized by \cite{Rasanen15} and extensively
discussed in \cite{Denissenya18}, with a heuristic suggestion that
appropriate weighting on the $\Omega_k$ estimator can provide a
substantially unbiased result. Therefore, in order to clarify and
study the systematics and scatter in our results, we will make
summary statistics in three ways: standard weighted mean, modified
weighted mean, and median statistics. Such a procedure has been
applied to the so-called $Om(z)$ diagnostic defined with the
expansion rate of the Universe \cite{Ding15b,Zheng16b}.

\begin{figure*}
\includegraphics[scale=0.6]{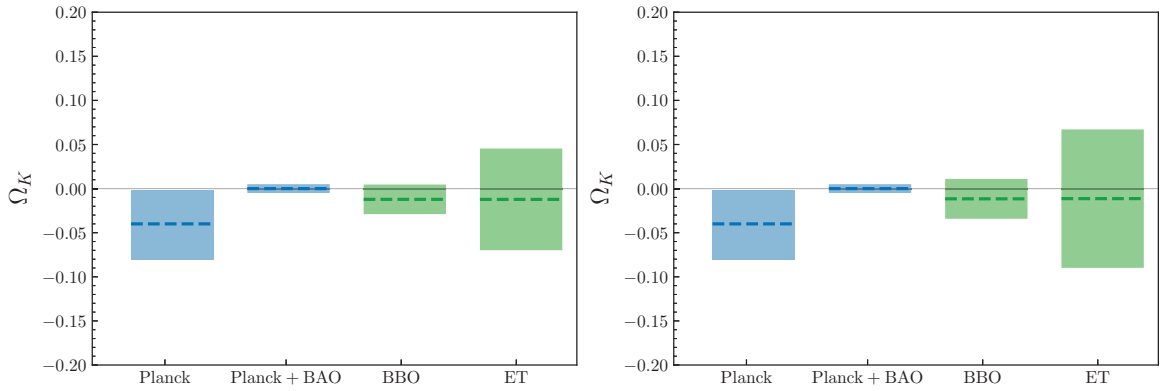}
\caption{Statistical summary of simulated predictions of the
$\Omega_k$ parameter measurements (inverse variance weighting) from
future observations of lensed GWs. Left and right panel respectively
show the results with 10\% and 50\% uncertainty in the amplification
factor measurements. Predictions for the ET and the BBO are
confronted with constraints achievable from the CMB and BAO
measurements.}\label{fig2}
\end{figure*}

The effectiveness of of our method could be seen from discussion of
the second question, that is: \emph{Is it possible to achieve a
stringent measurement of the present value of curvature density
parameter?} The most straightforward and popular way of summarizing
multiple measurements is inverse variance weighting:
\begin{equation}
\begin{array}{l}
\bar{\Omega_k}=\frac{\sum\left(\Omega_{k,i}/\sigma^2_{\Omega_{k,i}}\right)}{\sum1/\sigma^2_{\Omega_{k,i}}},\\
\sigma^2_{\bar{\Omega_{k}}}=\frac{1}{\sum1/\sigma^2_{\Omega_{k,i}}},
\end{array}
\end{equation} where $\bar{\Omega_{k}}$ stands for the weighted mean
of cosmic curvature and $\sigma_{\bar{\Omega_{k}}}$ is its
uncertainty. Let's start from the first case by introducing an
overall 10\% uncertainty to the amplification factor measurement,
concerning the estimation of luminosity distance to the lensed GW.
The forecast for the ET is: $\Omega_k=-0.011\pm0.057$. This result
illustrated in Fig.~\ref{fig2} is comparable to that derived from
the current estimation of the cosmic curvature
($\Omega_k=-0.040^{+0.038}_{-0.041}$) from the \textit{Planck} 2016
CMB data (the power spectra (TT, TE, EE+lowP)) \cite{Ade16}. Such
conclusion is also well consistent with the recent analysis of
\cite{Li19}, which discussed constraints on cosmological parameters
in FLRW metric by combining the time delay distances from lensed GW
signals, together with the co-moving distances obtained from a
parametrized fitting approach with independent SNe Ia observations.
In the case of the BBO providing more measurements of $\Omega_k$
with different redshift pairs, one can expect that cosmic curvature
could be estimated with the precision of $\Omega_k=-0.012\pm0.017$.
Admittedly, combination of CMB \textit{Planck} and BAO data leads to
a very high precision $\Omega_k=0.000\pm0.005$ but this result
pre-assumes the FLRW metric \cite{Ade16}. Such assumption can be
tested empirically by comparing the spatial curvature determined
from lensed GWs (detected by BBO) with that obtained from CMB+BAO
data. In the second case, when the fractional uncertainty of the
amplification factor measurement is assumed at the level of 50\%,
the resulting constraints on the cosmic curvature become
$\Omega_k=-0.012\pm0.073$ for the ET and $\Omega_k=-0.010\pm0.023$
for the BBO. The statistical results for $\Omega_k$ is shown in
Fig.~\ref{fig2}. From this plot it is evident that, in the framework
of the methodology proposed in this paper, the precision of derived
cosmic curvature is sensitive to the adopted amplification factor
measurements. This illustrates the importance of using auxiliary
data to improve constraints on the amplification factor, especially
the microlensing effect, with future more precise measurements for
local image environments and more knowledge on AGN accretion model
from astrophysics inputs \cite{Liao19}.

If one is to make summary statistics, one can do it in another two
ways. Firstly, considering the multiple powers of multiple distances
in the $\Omega_k$ estimation, different terms will prefer different
weighting \cite{Rasanen15}. Following the procedure proposed by
\cite{Denissenya18}, we apply purely empirical analysis to the
simulated sample and determine what weighting is most successful in
debiasing cosmic curvature. Our result show that appropriate
weighting with $1/\sigma_{\Omega_k}^{0.3}$ works well for $\Omega_k$
estimator. Proceeding this way, we have obtained
$\Omega_k=-0.006\pm0.092$ for the ET and $\Omega_k=-0.005\pm0.029$
for the BBO, by introducing an overall 10\% uncertainty to the
amplification factor measurement. Secondly, it is well known that
the weighted mean approach relies on several strong assumptions: a
statistical independence of the data, no systematic effects, and a
Gaussian distribution of the errors. Given the possible invalidity
of the above assumptions (especially the Gaussianity of errors),
another much more robust approach, the non-parametric median
statistics is also applied in our analysis, without the need to
assume anything about the error distribution. Such approach stems
from the well known property that for any particular measurement,
half of the data is expected to be higher and another half lower
than the median \cite{Gott01,Crandall13}. Therefore, the probability
that $n$-th observation out of the total number of $N$ is higher
than the median follows the binomial distribution:
$P=2^{-N}N!/[n!(N-n)!]$, which defines the $68.3\%$ confidence
intervals of the median \cite{Ding15b,Zheng16b}. In the framework of
the median statistics, when the fractional uncertainty of the
amplification factor measurement is assumed at the level of 10\%,
the resulting constraints on the cosmic curvature become
$\Omega_k=-0.007^{+0.018}_{-0.020}$ for the ET and
$\Omega_k=-0.006^{+0.016}_{-0.018}$ for the BBO. Therefore, one
should pay careful attention to the bias induced by nonlinearity of
the error propagation. For the current data sets of less than 1000
strong lens systems, this is of little real concern in the present
case, since the scatter always dominates over the bias in this case
\cite{Denissenya18}. If looking beyond this, one possible solution
to this issue can also be found in \cite{Rasanen15}, which applied
the $\chi^2$ statistics to fit constant $\Omega_k$ to 30
galatic-scale lenses from Sloan Lens ACS Survey. Their results
showed the lenses's goodness of fit could provide evidence for
deviations from the FLRW metric. If the FLRW hypothesis is not
rejected with reasonable value of $\chi^2/d.o.f.$, the probability
distribution of $\Omega_k$ could be directly obtained from the
$\chi^2$ values. For next generation data sets of more than 1000
strong lens systems, as was pointed in \cite{Denissenya18},
appropriate summary statistics only provide a demonstration of
principle that bias can be reduced. In this case, Monte Carlo
simulations of the actual data characteristics should be employed
for dealing with bias, which can be furthermore performed in an be
iterative way: subtract the modeled bias for $\Omega_k=0$, estimate
the new $\Omega_k$, and resimulate \cite{Denissenya18}.

To summarize, let us clarify some simplified assumptions underlying
our method. For instance, in this paper, we adopt the geometrical
optics approximation to derive the information of GW luminosity
distances, which is valid in all the observational events of
gravitational lensing of light. For the gravitational lensing of
gravitational waves, the wavelength is long so that the geometrical
optics approximation is not valid in some cases. As shown in
\cite{Takahashi03} and in a more detailed way in \cite{Zhang18}, the
wave optics should be used instead of the geometrical optics when
the wavelength of the gravitational waves $\lambda$ is longer than
the Schwarzschild radius of the lens mass $M_L$. More specifically,
the diffraction effect is important for $M_L\leq 10^8 M_\odot
(f/\rm{mHz})$ \cite{Takahashi03}. The wave effects become important
for the lens mass $10-10^4 M_\odot$ and $10^5-10^7 M_\odot$, which
is determined by the ET band ($10-10^4$ Hz) and the BBO band
($10^{-2}-1$ Hz), respectively. Therefore, in our approach this
effect does not significantly contribute to the scatter in the final
results. We make a final comment that, in order to implement our
method, dedicated observations including spectroscopic redshift
measurements of the lens and the source, velocity dispersion of the
lens, higher angular resolution imaging to measure the Einstein
radius, and dedicated campaigns to measure time delays would be
necessary. Obtaining these data for a large sample of strongly
lensed GWs will require substantial follow-up efforts. Despite of
these difficulties, the approach, introduced in this paper, might
provide a new window to engage multiple measurements of more
galactic-scale lensing systems where strongly lensed gravitational
waves and their electromagnetic counterparts can be simultaneously
detected. With a sample of lensed GWs, we could be optimistic about
detecting possible deviation from the FLRW metric within our
observational volume in the future. Such accurate measurements of
the FLRW metric can become a milestone in precision GW cosmology.

\section*{Methods}

\subsection*{Outline of the method}
We model the lens with a power-law mass distribution ($\rho \sim
r^{- \gamma}$) motivated by several previous studies, which found
that early-type galaxies are well described by power-law mass
distributions in regions covered by the X-ray and lensing
observations \cite{Humphrey10}, as well as the pixelated lens
potential corrections applied to gravitational lenses \cite{Suyu09}.
Based on the combination of the mass $M_{lens}$ inside the Einstein
radius and the dynamical mass inside the aperture $\theta_{ap}$
projected to lens plane, the spherical Jeans equation
\cite{Koopmans05} enables to assess the distance ratio
\cite{Cao12,Cao15}
\begin{equation} \label{Einstein}  \frac{d_{ls}}{d_s} = \frac{D^{A}_{ls}}{D^{A}_s} =   \frac{\theta_E} {4 \pi}
\frac{c^2}{\sigma_{ap}^2}  \left( \frac{\theta_E}{\theta_{ap}}
\right)^{\gamma-2} f(\gamma)^{-1}
\end{equation}
where $f(\gamma)$ is a certain function of the radial mass profile
slope \cite{Koopmans05} and $\sigma_{ap}$ is the luminosity
averaged line-of-sight velocity dispersion of the lens inside the
aperture. It is clear that $\sigma_{ap}$, $\theta_E$, $\theta_{ap}$
and $\gamma$ obtained from the observations enable one to measure
the distance ratio $d_{ls}/d_s$.

Time delays between GW signals $\boldsymbol{\theta}_i$ and
$\boldsymbol{\theta}_j$ depend on the "time-delay distance"
($D_{\mathrm{\Delta t}}$) and the lens mass distribution
\cite{Treu10}
\begin{equation}
\Delta t_{i,j} = \frac{D_{\mathrm{\Delta
t}}(1+z_{\mathrm{l}})}{c}\Delta \phi_{i,j}, \label{relation}
\end{equation}
where
$\Delta\phi_{i,j}=[(\boldsymbol{\theta}_i-\boldsymbol{\beta})^2/2-\psi(\boldsymbol{\theta}_i)-(\boldsymbol{\theta}_j-\boldsymbol{\beta})^2/2+\psi(\boldsymbol{\theta}_j)]$
is the Fermat potential difference determined by the lens
mass distribution and the source position $\boldsymbol{\beta}$,
$\psi$ denotes two-dimensional lensing potential determined by
the Poisson equation $\nabla^2\psi=2\kappa$, where $\kappa$ is
dimensionless surface mass density (convergence) of the lens. Hence, the time-delay distance is
\begin{equation}
D_{\mathrm{\Delta t}}\equiv\frac{D^A_{\mathrm{l}}
D^A_{\mathrm{s}}}{D^A_{\mathrm{ls}}}=\frac{c}{1+z_l}\frac{\Delta t_{i,j}}{\Delta \phi_{i,j}}.
\end{equation}
Therefore, the combined observations of strongly lensed EM and GW
signals coming from the same source will provide
the comoving distance to the lensing system
\begin{equation}
D_l=(1+z_l)D_{\mathrm{\Delta t}} \frac{D^{A}_{ls}}{D^{A}_s}.
\end{equation}


Inspiralling compact binary systems act as self calibrating standard
sirens providing the luminosity distance to the source. However,
there is a degeneracy between GW inferred luminosity distance $D^L$
and lensing magnification $A \sim \sqrt{\mu} / D^L$. Detailed
analysis of the images of host galaxy and accompanying EM
counterpart, can provide a measurement of the magnification of
images and break down the degeneracy in GW based inference. It can
be best illustrated with the singular isothermal sphere (SIS) lens,
when two images $x_{\pm} = y \pm 1$ are formed ($x=\theta /
\theta_E$, $y = \beta / \theta_E$) with flux magnification
$\mu_{\pm} = 1/y \pm 1$. From the flux ratio one can infer the
source position $y = \frac{1-F_{-}/F_{+}}{1 + F_{-}/F_{+}}$. Once
the source position is known, one can quantify the amplification of
each image and estimate the luminosity distance $D^L$ from GW
waveforms. Therefore, the transverse comoving distance from the
observer to the source could be derived as
\begin{equation}
D_s= \frac{1}{1+z_s} D^L_s
\end{equation}
provided the GW source's redshift is known from the identification
of an EM counterpart and its host galaxy. Combining the above
analysis with Eq.~(2), one can see that the function $\Omega_k(z_l,
z_s)$ can be directly obtained.


\subsection*{Simulated lenses}

We conservatively consider only elliptical galaxies, which
contribute $\sim 80\%$ to the total lensing probability
\cite{Oguri10}. It was found  that grid-based lens potential
corrections from power-law models were only 2\% \cite{Suyu09},
further justifying the use of a simple power-law model to describe
the mass distribution even for complicated lenses. In particular,
various studies have shown that the power-law profile provides an
accurate description of lens galaxies, out to $z_l\sim1$, which are
observed in number of large surveys (see Ref.~\cite{Cao15} for the
lens redshift distribution in SLACS, BELLS, SL2S and LSD). At this
point, one should clarify the issue whether the power-law model is
valid for high-redshift lenses, since it was suggested that LSST is
also capable of discovering higher redshift lenses than currently
known \cite{Collett15} (the lens redshift of our simulated sample
may reach to $z_l\sim1.9$). However, in our analysis, the magnitude
of uncertainty generated by such issue might be overestimated: on
the one hand, the well-known modified Schechter function
\cite{Sheth03} already predicts no significant lens population at
high redshift; on the other hand, although the high-redshift
galaxies with measured velocity dispersions is small, no significant
dependence of the slope parameter $\gamma$ on redshift (in the
framework of power-law model) has been found so far based on lensing
and dynamical analysis \cite{Koopmans09}. In order to ascertain
similarity between our simulations and the real world, we assumed
velocity dispersion of lenses as $\sigma_{ap}=210\pm50$ km/s and
lens redshift distribution with the median value of $z_l=0.8$, which
is consistent with the properties of the LSD sample. We are thus
confident that the simulated population of lenses is a good
representation of what the future surveys might yield, considering
the similarity of the redshift distribution of discoverable lenses
in forthcoming LSST survey with that found in \cite{Gavazzi14}.

\subsection*{Lensed GW mock catalog}

One can expect that both ET and BBO should register a considerable
catalog of such events during a few years of successful operation:
from 5 $\sim$ 10 years' accumulated data $\sim$ 100 lensed GW events
will be detected by three nested ET interferometers in the redshift
range $z \leq 5.00$ \cite{Liaokai2017,Li18}. The BBO, a proposed
space-based GW detector, would possibly detect $\sim$ 1000 lensed GW
events from $10^6$ compact-star binaries \cite{Cutler09}.
Construction of the mock catalog proceeded along the following
steps. The lensing rate strongly depends on the estimate of the GW
event rate, which depends on the estimate of the merger rate of
double compact objects (DCO). In this paper, we have adopted the
conservative SFR function from \cite{Dominik13} and taken the data
from the website \emph{http: www.syntheticuniverse.org}, the
so-called "rest frame rates" in cosmological scenario. Concerning
gravitational lensing, the velocity dispersion distribution in the
population of early-type galaxies was modelled as modified Schechter
function with parameters from the SDSS DR3 data \cite{Choi07}. Based
on the intrinsic merger rate of DCO calibrated by strong lensing
effects \cite{Ding15}, we obtained the differential rates of lensed
GW events as a function of $z_s$, which furthermore constituted the
sampling distribution of lensed GWs.

For each lensed GW event, the mass of the neutron star, the mass of
the black hole, and the position angle $\theta$ are randomly sampled
in the three parameter intervals: $[1,2] \; M_\odot$, $ [3,10] \;
M_\odot$, and $[0,\pi]$ \cite{Cai15}. It was well acknowledged that
luminosity distances can be inferred from the waveform of GW signals
from chirping binaries. Different sources of uncertainties are
included in our simulation of luminosity distance $D^L_s$. Firstly,
for ET, the combined SNR for the network of three independent
interferometers is defined by the inner product \cite{Cai17} of the
Fourier transform $H(f)$ of the time domain waveform $h(t)$ , which
not only confirms the detection of GW with $\rho_{net}>8.0$, but
also contributes to the error on the luminosity distance as
$\sigma_{inst}\simeq 2D^L_s/\rho_{net}$. In our calculation, the
upper cutoff frequency is dictated by the last stable orbit
$f_{upper} = 2f_{LSO}$, while the lower cutoff frequency is taken as
$f_{lower} = 1 Hz$. The BBO, which is fundamentally
self-calibrating, would determine the luminosity distance to each
binary to 1\% accuracy. More specifically, we assume that the
distance measurement errors due to detector noise for each
individual binary are those shown in \cite{Cutler09}. Secondly,
following the strategy described by \cite{Sathyaprakash10}, weak
lensing has been estimated as a major source of error on $D^L_s(z)$
for standard sirens. For the ET we estimated the uncertainty from
weak lensing according to the fitting formula \cite{Zhao11}
$\sigma_{lens}/D^L_s=0.05z$ . For the BBO, since the systematic
distance errors arising from the detector itself will also be
negligible, the uncertainty of the true distance to the binary
system will be dominated by the effects of weak lensing
\cite{Cutler09}, parameterized as $\sigma_{lens}/D^L_s=0.044z$.

In the case of strongly lensed GWs, the luminosity distance could be
estimated with a better knowledge of the lensing amplification
factor for the GW signal. More specifically, the lensing
amplification factor (of the GW signal) could be derived from the
lensing magnification (of the EM observation), with the latter
determined by solving the lens equation using \emph{glafic}
\cite{Oguri10b}. Therefore, the uncertainty of $F$ is related to
that of the lens mass profile and the Einstein radius, which can be
derived from high-quality imaging observations \cite{Suyu13}. All
lensed GW signals (images) will be used to determine luminosity
distances and corresponding lensing amplification factors. Note that
the lensing magnification is dependent on the lens model, which is
the only way to determine the magnification for almost all strong
lensing systems such as the quasar-galaxy and galaxy-galaxy systems
\cite{Oguri10,Oguri10b}. We emphasize that only for the strongly
lensed standard candles, SNe Ia, the magnification can be derived
from comparing the observed brightness to other SNe Ia within a
narrow redshift range \cite{Goobar17}, such measurement of the
lensing magnification is independent of any assumptions on cosmology
and lens model. Finally, the influence of the microlensing (ML)
effect generated by stars in lensing galaxy should be considered in
the estimation of amplification factors. Due to the inclination of
the finite AGN accretion disc and the differential magnification of
the coherent temperature fluctuations, the microlensing by the stars
can lead to changes in the actual magnification of the lensed EM
signal (e.g. flux-ratio anomalies). The problem was recently
recognized concerning galactic-scale strong lensing systems with SNe
Ia as background sources, with a heuristic suggestion that adding an
additional uncertainty $\sim 0.70$ mag to lensed SNe Ia
\cite{Yahalomi17}. However, there still exist a lot of uncertain
inputs for the microlensing amplification factor priors, concerning
thehost galaxy or accompanying EM counterpart of GW. On the one
hand, it has been recently proved that the adopted local
environments for images and mass function for the stars, which can
generate different local convergency, shear and star proportion,
could systematically bias the magnification map \cite{Chen18}. On
the other hand, the inclination, position angle, especially the size
of the accretion disk, as well as the relative motion of the source,
would also bring uncertainties \cite{Tie18}. Recent discussion on
this issue can be found in \cite{Liao19}. Taking the above factors
into consideration, in our simulated data we choose to respectively
introduce 20\% and 100\% uncertainty to the magnification
measurements, which corresponds to 10\% and 50\% uncertainty to the
amplification factor ($\delta F=10\%, 50\%$), and quantify its
effect on the cosmic curvature estimation.

\subsection*{Cosmic curvature estimation}

Now one is able to determine the cosmic curvature as
\begin{equation} \label{k}
\Omega_k (z_l, z_s) = \frac{d_l^4 + d_s^4 + d_{ls}^4 - 2 d_l^2 d_s^2
- 2 d_l^2 d_{ls}^2 - 2 d_s^2 d_{ls}^2}{4 d_l^2 d_s^2 d_{ls}^2}
\end{equation}
It should be noted that, when using the distance information of the
lensing system itself, one does expect errors from $d_l$ to be
covariant with $d_{ls}$ and $d_s$. In particular,
\cite{Denissenya18} addressed the problem of distance correlation
and proposed using a different combination of distances ($d_{\Delta
t}$, $d_l$ and $d_s$) to get $\Omega_k$. The dimensionless
time-delay distance is related to the time-delay distance as
$D_{\Delta t}=cd_{\Delta t}/H_0$. Because distances measured at
widely separated redshifts should be mostly uncorrelated, they
proposed that $d_l$ and $d_s$ could be derived from the
reconstruction of the distance-redshift relation in a model
independent manner, concerning the available standardized candle
(Type Ia supernovae) or standard ruler (BAO) data. Putting together
Eqs.~(\ref{Einstein})-(\ref{k}), one can rewrite the cosmic
curvature in terms of measurable distances
\begin{equation} \label{k2}
\Omega_k (z_l, z_s) = \frac{1}{4}\frac{(1+z_l)^2d_{\Delta
t}^2}{d_s^4}+\frac{1}{4}\frac{(1+z_l)^2}{d_{\Delta t}^2
(d_{ls}/d_s)^4}+\frac{1}{4}\frac{1}{(1+z_l)^2d_{\Delta
t}^2}-\frac{1}{2}\frac{1}{(d_{ls}/d_s)^2d_s^2}-\frac{1}{2}\frac{1}{d_s^2}-\frac{1}{2}\frac{1}{(1+z_l)^2d_{\Delta
t}^2(d_{ls}/d_s)^2}
\end{equation}
which will be the central equation in our analysis. In order to
carry out the curvature estimation, we need a measurement of
$d_{ls}/d_s$ from image configuration, lens modeling and dynamics,
$d_{\Delta t}$ from a strongly lensed time delay system, and $d_s$
from the lensed standard siren. In this paper, instead of
propagating distance uncertainties, we turn to the other effective
solution to this problem, i.e., the uncertainty on the determination
of the curvature is given in terms of the measurement uncertainties
on the observables characterizing the lens mass profile ($\gamma$,
$\theta_E$, $\theta_{ap}$), time delays ($\Delta t$), Fermat
potential difference due to lens and line-of-sight effect
($\Delta\psi$, $\Delta\psi$ (LOS)), the luminosity distance to the
source ($d_s^L$), and the amplification factor ($F$ with
microlensing effect). Note that in the formalism presented in this
paper, the mass profile of the lensing galaxy is described by
$\gamma$ and $\theta_E$, which is related to the lens potential
difference ($\Delta \psi$). The amplification factor also depends on
the lens model, and thus the mass profile of the lensing galaxy.
However, considering the uncertain influence of the microlensing
(ML) effect, in our analysis we choose to introduce an overall 10\%
and 50\% uncertainty to the amplification factor measurements. The
uncertainties in these parameters are presumably covariant, which
indicates that per-parameter uncertainties treated independently
would be expected to underestimate propagated uncertainties. After
considering such covariance, the uncorrelated measurement
uncertainties on the observables of $\gamma$, $\theta_E$,
$\theta_{ap}$, $\Delta t$, $\Delta\psi$ (LOS), $d_s^L$, and $F$ will
be projected on the determination of $\Omega_k$ (Eq.~(\ref{k2}))
\begin{equation} \label{k3}
\delta \Omega_k (z_l, z_s) \sim ( \delta \gamma, \delta \theta_E,
\delta \theta_{ap}, \delta (\Delta t), \delta (\Delta\psi (LOS)),
\delta d_s^L, \delta F).
\end{equation}
More specifically, given the nonlinear combination of observables
that goes into the curvature estimation, we use Monte-Carlo
simulation to project uncertainties onto the final uncertainty of
$\Omega_k (z_l, z_s)$.


\section*{Acknowledgements}

This work was supported by National Key R\&D Program of China No.
2017YFA0402600, the National Natural Science Foundation of China
under Grants Nos. 11503001, 11690023, 11373014, and 11633001, the
Strategic Priority Research Program of the Chinese Academy of
Sciences, Grant No. XDB23000000, the Interdiscipline Research Funds
of Beijing Normal University, and the Opening Project of Key
Laboratory of Computational Astrophysics, National Astronomical
Observatories, Chinese Academy of Sciences. J.-Z. Qi was supported
by China Postdoctoral Science Foundation under Grant No.
2017M620661. M. Biesiada was supported by Foreign Talent Introducing
Project and Special Fund Support of Foreign Knowledge Introducing
Project in China. He is also grateful for support from Polish
Ministry of Science and Higher Education through the grant
DIR/WK/2018/12. Y. Pan was supported by the Scientific and
Technological Research Program of Chongqing Municipal Education
Commission (Grant no. KJ1500414); and Chongqing Municipal Science
and Technology Commission Fund (cstc2015jcyjA00044, and
cstc2018jcyjAX0192).

\section*{Author contributions statement}

S. Cao contributed in proposing the idea using strongly
gravitational lensed gravitational waves to directly test the FLRW
metric. He also contributed in original paper writing. J.-Z. Qi
contributed in proposing the measurement strategy and calculations.
M. Biesiada and J. Li contributed in data simulations and improving
the quality of the paper. Y. Pan contributed in comparing the work
with relevant literature. Z.-J. Cao and Z.-H. Zhu contributed in
discussing the new ideas and organizing the research.

\section*{Additional information}

The authors declare no competing interests.

\newpage

\begin{table*}
\begin{center}
\begin{tabular}{lccc}
\hline\hline

& $\delta\theta_E$ & $\delta\sigma_{ap}$ & $\delta\gamma$ \\
\hline
Image configuration   &1\% & 5\% & 1\%\\
\hline
& $\delta\Delta t$ & $\delta\Delta\psi$ &  $\delta\Delta\psi (LOS)$  \\
\hline
Time delay  &0\% & $\sim (\delta\theta_E,\delta \gamma)$ &1\% \\
\hline & $\delta d^L_s$ (SNR) & $\delta d^L_s$ (WL)  & $\delta F$ (SL+ML)  \\
\hline
Lensed GW & $ 2/\rho_{net}$ & $0.05z$   & 10\% (50\%) \\
\hline\hline
\end{tabular}
\end{center}
\caption{Relative uncertainties of respective factors contributing
to the accuracy of $\Omega_k(z_l,z_s)$ measurement.
$\delta\theta_E$, $\delta\sigma_{ap}$, and $\delta\gamma$ denote
Einstein radius, aperture velocity dispersion, and mass-density
profile; $\delta\Delta t$, $\delta\Delta\psi$, $\delta\Delta\psi
(LOS)$ correspond to time delay, Fermat potential difference and
light-of-sight contamination, respectively; $\delta d^L_s$ and
$\delta F$ correspond to the dimensionless luminosity distance and
amplification factor of lensed GW. }\label{error}
\end{table*}

\end{document}